\documentstyle[aps,preprint]{revtex}

\begin{document}

\draft

\title{Signature of metastable electrons in highly charged ion surface
  interactions}

\author{J.~Ducr\'ee\thanks{Author to whom correspondence should be
    addressed. Electronic address: ducree@uni-muenster.de},
  J.~Mrogenda, E.~Reckels, M.~R\"uther, A.~Heinen, Ch.~Vitt,
  M.~Venier, J.~Leuker, and H.J.~Andr\"a}
  
\address{Institut f\"ur Kernphysik, Westf\"alische
  Wilhelms-Universit\"at M\"unster, Wilhelm-Klemm-Str.~9, D-48149
  M\"unster, Germany}

\date{\today}

\maketitle

\begin{abstract}
   
  We present autoionization spectra of metastable Ar$^{8+}$ and
  C$^{4+}$, N$^{5+}$, O$^{6+}$ and Ne$^{8+}$ scattering off an Al(111)
  surface with incident energies down to 5~eV. The unprecedented
  quality of the experimental data permits the observation of a unique
  in the structures originating from the metastable projectiles
  compared to corresponding ground state configurations.  Analyzing
  the peak positions for different projectile species and velocities
  we demonstrate that the peak must be ascribed to an above-surface
  transition under participation of the metastable state.
\end{abstract}


The interactions of highly charged ions (HCIs) with surfaces have been
studied extensively in the past, refer to \cite{Arn97} for a recent
review. Several research groups have collected a tremendous amount of
experimental data on Auger emission~\cite{Mey91,Gre96,Tho97}, X-ray
emission \cite{And91A,Bri96} and energy gains of reflected projectiles
\cite{Win96A}.  Also theoretical analysis has reached a remarkable
level of sophistication \cite{Die96,Lim95a}.  Nevertheless, authors
still argue about crucial issues of the interaction mechanism since a
lot of the interpretations deduced from the experiment depend on
certain model assumptions.

As a HCI approaches a metal surface it is neutralized via resonant
charge transfer from the target valence band (VB) into its outer
Rydberg states, see Fig.~\ref{fig:model}. Due to the image attraction,
the interaction time of the incoming projectile in front of the
surface possesses a principal lower limit of the order of
$10^{-14}$~s.  Close above the first bulk layer atomic orbitals are
energetically shifted and strongly distorted due to the bulk
interaction and a theoretical description becomes intricate.

Authors have concluded from atomic structure calculations neglecting
the bulk perturbations that a stepwise intra-atomic autoionization
cascade proceeds too slow in order to carry all Rydberg state
electrons into inner shells in front of the surface. Indeed, striking
experimental evidence has been provided indicating that the
overwhelming fraction of the inner-shell processes, in particular {\it
  KLL\/} emission, takes place within the bulk region
\cite{Lim95a,Koe94,Gre95}.

In this domain, the interactions between HCIs and the metal electron
gas are fundamentally different. As the neutralized HCI reaches the
first bulk layer it experiences the screening of the high density
target electron gas and outer levels merge into the VB\@.  Only
electrons on the innermost projectile orbitals have a chance to
survive bulk penetration. The HCI now acts as a point charge
attracting a VB charge cloud to screen the positive core charge.

Since the image charge components and thus the potential barrier
\cite{Duc97} which builds up between the HCI and the surface vanishes
within the bulk the charge cloud electrons can directly replenish
inner projectile levels via so called $LCV$ transitions
\cite{Die96,Duc97A}, see also Fig.~\ref{fig:model}. Compared with the
slow intra-atomic cascade above the surface, this filling mechanism
and succeeding Coster-Kronig regrouping \cite{Sch94} of the $L$-shell
population proceed very rapidly and higher initial sublevel
populations evolve before an inner shell transition takes place.

Auger transition energies increase significantly when inner-shell
populations shield the core in a way that the statistically preferred
shell configurations at the time of $KLL$ decay can be reconstructed
via the most prominent spectral peak positions. Various efforts have
quite successfully been conducted to develop an appropriate
interaction model from these experimental clues combined with
theoretical work \cite{Gre95,Sch94}.

In this paper we address the central issue of above-surface
contributions in a new manner. In contrast to previous efforts in this
field which partially rely on rather complex model assumptions
concerning the evolution of the filling cascade \cite{Lim95a,Das93} or
secondary electron transport through the bulk region \cite{Koe94}, the
discovery of a new spectral structure allows us to only make use of
simple, commonly accepted concepts.

\section{Experimental Spectra}
\label{sec:spectra}

The HCIs are extracted by a fixed voltage of --20kV from an ECR ion
source floated on a selectable potential $U_{\mbox{\scriptsize S}}$.
After the ions are $q/m$ separated, a lens system focuses and
decelerates the beam onto the grounded target in a UHV chamber ($p
\simeq 5 \times 10^{-12}$hPa).  An ion spectrometer, mounted close
behind the removable target, measures the kinetic ion energy
distribution the full width of which at half maximum never exceeded
2~eV per charge.  The beam energy $E_{\mbox{kin}}$ can freely be set
by varying $U_{\mbox{\scriptsize S}}$.  The beam axis intersects the
target surface at a variable angle $\Theta$.  Electrons are detected
perpendicular to the beam axis at an angle of $\Psi=90^\circ-\Theta$
with respect to the surface by an electrostatic 150$^\circ$ spherical
sector analyzer. The Al(111) surface has been prepared by successive
cycles of sputtering and annealing and controlled by Auger electron
spectroscopy.

In Fig.~\ref{fig:argon} we display two spectra of Ar$^{8+}$ and
Ar$^{9+}$ projectiles impinging with $E_{\mbox{kin}}=12.5$~eV per
charge and $\Theta=5^\circ$ on an Al(111) surface. All spectra in this
paper are normalized to the area below the portrayed peak region. The
comparison with the Ar$^{9+}$ spectrum yields a distinct, yet unseen
peak in the well resolved Ar$^{8+}$ spectrum on the upper fringe of
the {\it LMM\/} region which overlaps the $LMN$ region. Furthermore, a
notable fraction of the ($3p \mapsto 2p, 3p \uparrow$) intensity in
the dominant 211-eV peak shifts to the ($3s \mapsto 2p, 3\ell
\uparrow$) region ($\ell \in \{s,p\}$) at lower energies in the
Ar$^{8+}$ spectrum as a consequence of the additional $3s$ electron.
For higher perpendicular incident velocities at $\Theta=45^\circ$ the
247-eV peak vanishes. Other aspects of these Ar$^{q+}$ spectra have
been outlined in \cite{Duc97A}.

In Fig.~\ref{fig:series} we recorded a series of spectra of HCIs in
metastable $1s2s$ and hydrogenlike $1s$ ground state configurations
interacting with an Al(111) surface at $\Theta=5^\circ$ with
$E_{\mbox{kin}}=14.5$~eV per charge.  On the analogy of the Ar$^{8+}$
spectrum in Fig.~\ref{fig:argon}, the additional peak (marked by
vertical lines in the plots) within the $1s2s$ spectra is superposed
on the high-energy tail of the $KLM$ region.  It is most prominent for
C$^{4+}$ and widens along the row towards Ne$^{8+}$. For C$^{4+}$ we
added a spectrum measured at $E_{\mbox{kin}}=5$~eV and
$\Theta=5^\circ$ on Si(100) to demonstrate that the height and
sharpness of the additional peak can be enhanced by reducing the
incident velocity. In other measurements on Si(100) we discovered
nearly identical $KLV_W$ regions as on Al(111).

The greatest intensity of the spectra in Fig.~\ref{fig:series}
originates from $KLL$ transitions. The two sharp peaks (a and b) on
the low-energy side have previously \cite{And91A,Lim95a} been assigned
to ($2s \mapsto 1s,2s \uparrow$) and ($2s \mapsto 1s,2p \uparrow$)
transitions out of low $L$-shell occupations. The upper part of the
spectrum is partially composed of a ($2p \mapsto 1s,2p \uparrow$)
region (c) featuring a less pronounced peak on the high-energy side.
With increasing effective core charge towards Ne$^{8+}$ a well
separated $KLM$ peak (d) emerges.

Fig.~\ref{fig:series} unveils further striking discrepancies in the
overall $1s2s$ and $1s$ peak structures. For the first time, we
identify a systematic expansion of the ($2s \mapsto 1s, 2s \uparrow$)
peak (a) to the low-energy side for metastable projectiles which is
most pronounced for Ne$^{8+}$. In a tentative explanation, we ascribe
this remarkable shift to the close HCI-surface encounter region where
the projectile core is still incompletely screened by the target
electron gas while the stripping of outer Rydberg levels has already
set in.  Within the isotropic bulk environment, level shifts only
depend on the spatially constant screening of valence band electrons
and sharp peak profiles like the 211-eV peak in Fig.~\ref{fig:argon}
occur. We postpone the discussion of this major effect to a future
publication. Additionally, well defined subpeaks in the ($2p \mapsto
1s, 2p \uparrow$) (c) and a $KLM$ region (d) can be resolved.
Complying with the previous peak assignment, regions (b) and (c)
exhibit less intensity with respect to region (a) in the spectra of
metastable projectiles.

\section{Interpretation of the peak position}
\label{sec:interpretation}

For the spectra of metastable configurations, the peculiar position of
the additional peak on the high-energy tail of the $KLM$ spectra
implies that apart from the $2s$ electron which jumps into the
$1s$ vacancy an energy level situated far above the $M$-shell has to
supply the emitted electron. In an attempt to find a consistent
explanation for all HCI species we found that the peak position can
most accurately be reproduced by putting the emitted electron in a
projectile energy level which is resonant to the target work function
$W$. The evaluation of the transition energy has been performed by the
Cowan code~\cite{Cow81} selecting $1s2sns^xnp^y$ as initial and
$1s^2ns^xnp^{y-1}$ as final configurations. The initial configuration
is made neutral by choosing $x+y=q-2$ where $q$ denotes the nuclear
projectile charge. $n$ represents the principal quantum number the
$p$-subshell binding energy of which comes closest to $W$.

In Table~\ref{tab:KLVW} we list the initial shell populations, the
binding energy of the outermost $np$-subshell, which should be
compared to the Al work function of 4.25~eV, and the experimental and
calculated peak energies. For each HCI species, including Ar$^{8+}$
for which we replaced the core configuration $1s2s$ by $2p^53s$, we
attain surprisingly good agreement with the experimental values
despite the simple evaluation method. In the following we denote these
transitions by $KLV_W$ and for Ar$^{8+}$ by $LMV_W$ and summarize the
$n$-shells which are nearly resonant to $W$ by $V_W$.  We now have to
tackle the problem where this unique transition type takes place. The
exclusive occurrence of the $KLV_W$ peak at low perpendicular incident
velocities and its weak intensity suggest that it might proceed during
the limited interaction time in front of the bulk.  Further evidence
for this hypothesis comes from the scrutiny of the effective
potentials seen by an active electron in front of and underneath the
surface in the oncoming paragraphs.

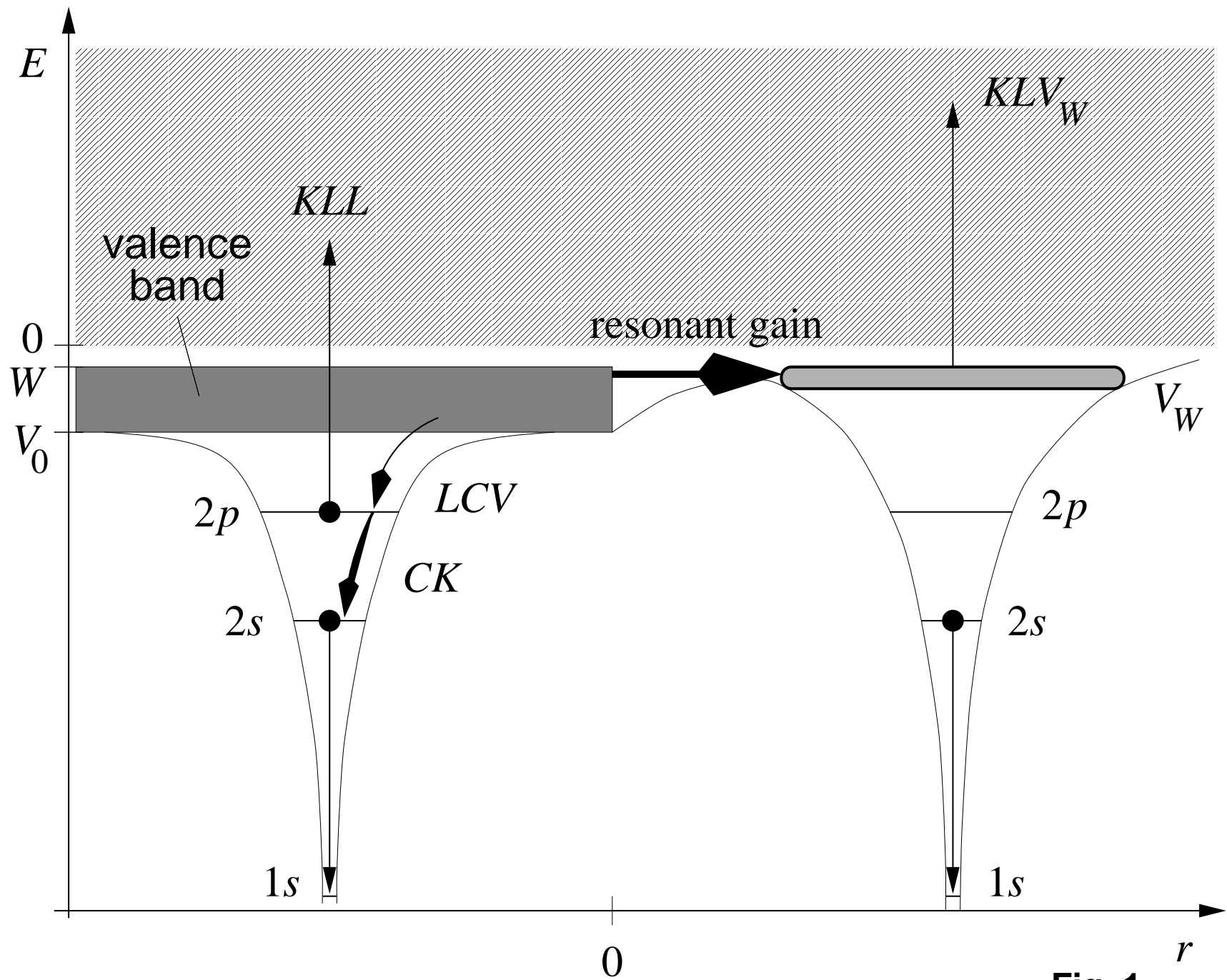
\begin{figure}[p]
  \caption{Sketch of the interaction scenario. Above the surface
    valence band electrons are resonantly transferred across the
    potential barrier into HCI Rydberg levels which accumulate large
    populations due to the slow intra-atomic cascade. The energy of
    the resonant Rydberg levels which we globally denote by $V_W$
    remains constant at the target work function $W$ while quantum
    numbers may change.  For metastable projectiles the $2s$ electron
    can participate in a so called $KLV_W$ transition creating
    a peak which cannot occur during below surface emission. An
    analogous argumentation holds for the $LMV_W$ peak in
    Ar$^{8+}$ spectra.}
  \label{fig:model}
\end{figure}

The right hand side of Fig.~\ref{fig:model} schematically depicts the
deformation of the projectile potential close to the surface.  In the
classical overbarrier model \cite{Bur91} the charge transfer is
described via one-electron Coulomb and image potentials. A narrow
channel of typically a few eV width \cite{Duc97} classically permits a
resonant charge flow between the HCI and the metal VB\@. The upper
boundary of the channel is given by the highest occupied VB states,
i.e.\ the work function $W$. The resonant capture is believed to
constitute the most efficient above-surface neutralization mechanism.

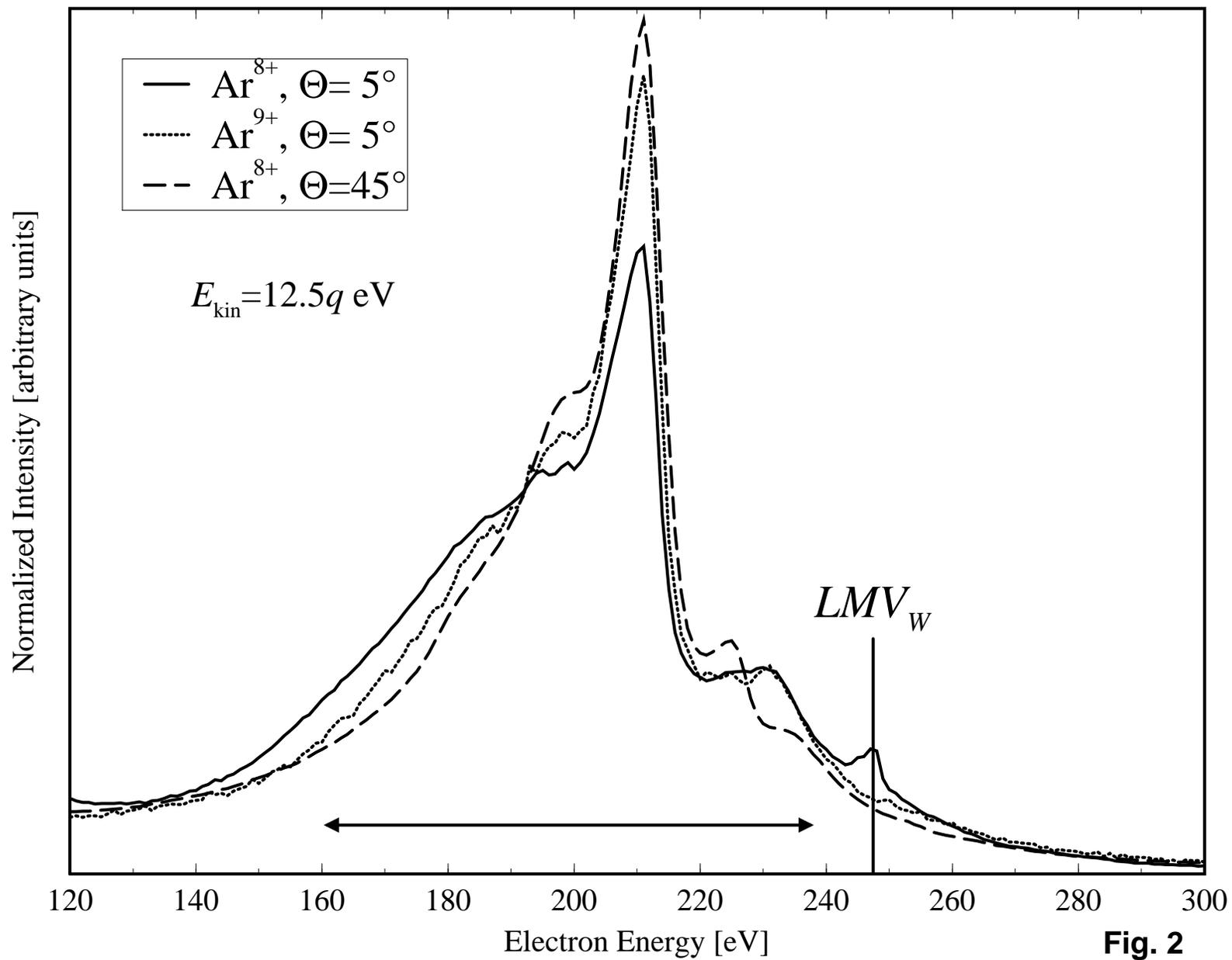
\begin{figure}[p]
  \caption{Spectra of  Ar$^{9+}$ and metastable Ar$^{8+}$ impinging
    with $E_{\mbox{kin}} = 12.5q$~eV at incident angles
    $\Theta=5^\circ$ and $45^\circ$ on Al(111), featuring an as yet
    undiscovered peak above the $LMM$ region for Ar$^{8+}$ under
    the lowest perpendicular incident energy. At the $LMV_W$
    peak, the absolute count rate amounts to 23000 electrons per 1~eV
    channel and the statistical error is well below 1\%. We ascribe
    this peak to $LMV_W$ processes, cf.\ 
    Fig.~\protect{\ref{fig:model}}.}
  \label{fig:argon}
\end{figure}

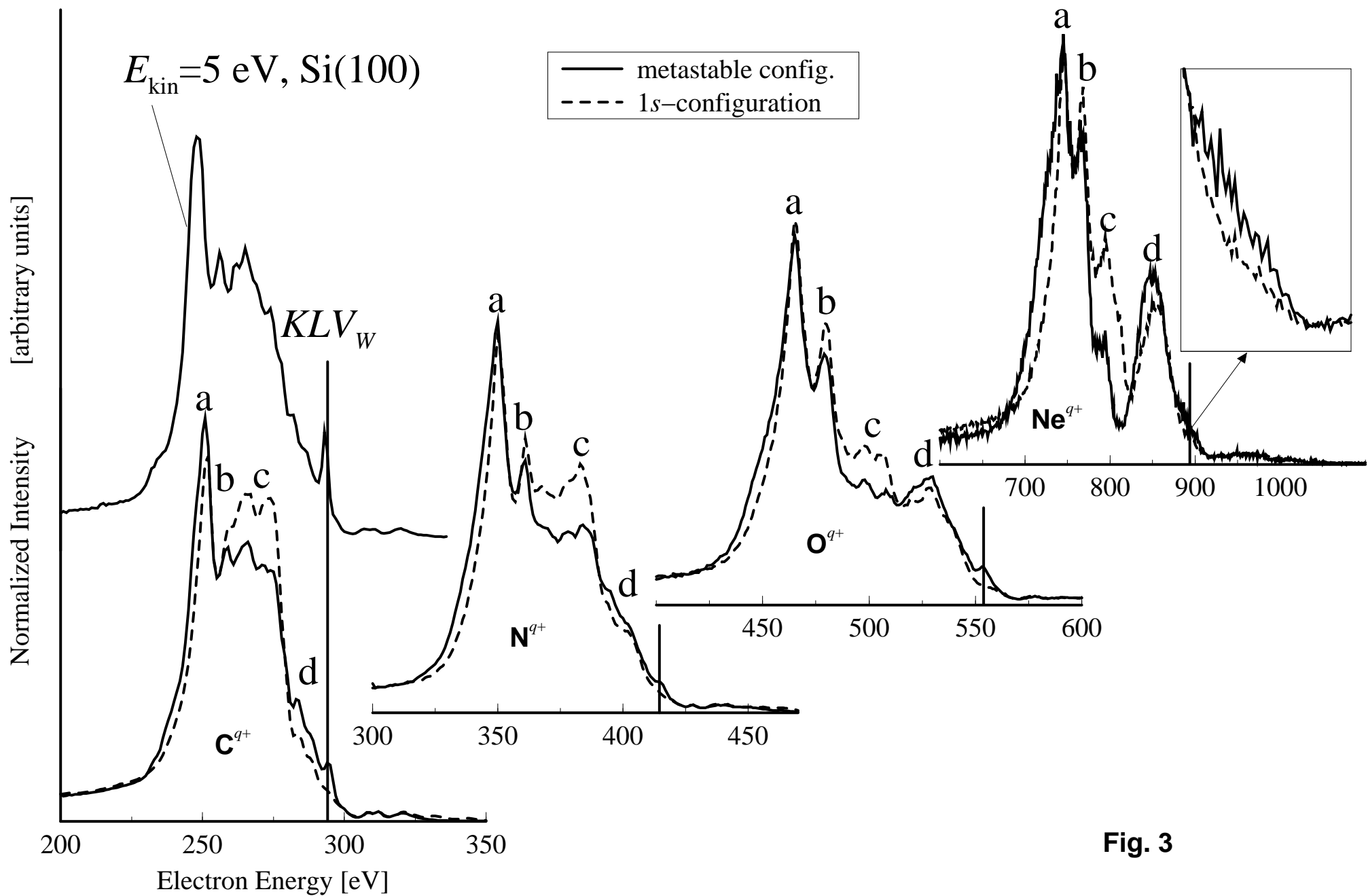
\begin{figure}[p]
  \caption{Spectral series of metastable ($1s2s$) HCIs scattering
    off an Al(111) surface at $\Theta=5^\circ$ $E_{\mbox{kin}}$ and
    very low perpendicular incident velocities at $E_{\mbox{kin}} =
    14.5q$~eV and for C$^{4+}$ also at 5~eV on Si(100) with
    $W_{\mbox{Si}}=4.6$~eV. a, b, c and d label the ($2s \mapsto 1s,
    2s \uparrow$), ($2s \mapsto 1s, 2p \uparrow$), ($2p \mapsto 1s, 2p
    \uparrow$) and $KLM$ regions which are well separated in the
    Ne$^{9+}$ spectrum, respectively.  The comparison with the
    corresponding $1s$ configuration spectra reveals an additional
    $KLV_W$ peak (cf.\ Fig.~\protect{\ref{fig:model}}) above
    the $KLM$ region (marked by a vertical line) which is most
    pronounced for C$^{4+}$.  At a typical count rate of 96000
    electrons per 1-eV channel at the C$^{4+}$ $KLV_W$ maximum
    the experimental error falls far below 1\%.}
  \label{fig:series}
\end{figure}

For metastable projectiles, the resulting atomic electron
configuration complies with the Cowan code simulation in
Tab.~\ref{tab:KLVW} and represents the ideal system to generate the
$LMV_W$ and $KLV_W$ peaks in Figs.~\ref{fig:argon}
and~\ref{fig:series}. The sharpness of the $KLV_W$ peak for metastable
C$^{4+}$, N$^{5+}$ and the $LMV_W$ peak of Ar$^{8+}$ suggests that
captured electrons stay parked in the $V_W$ levels during the
above-surface interaction phase which would be in contradiction to the
mere cascade model \cite{Das93}. We assume that this behavior results
from the dynamic interplay of resonant capture and loss and the
elevation of projectile $V_W$ levels by image shifts. It is important
to note these shifts vanish in commonly employed models \cite{Duc97}
for the neutralized projectiles in Tab.~\ref{tab:KLVW}. For the case of
ionic projectiles, the continuum boundary limits the upward shifts of
resonantly populated levels to less than $W$, i.e.\ a few eV only.
Since Auger transition rates between two subshells critically depend
on the energetic vicinity of the participating levels, high {\it
  KLV}$_W$ peak intensities cannot be expected despite the massive
$V_W$ level occupation.

Within the bulk, the screening of the target electron gas lets the
atomic potential merge into $V_0$ at large distances from the nucleus.
$LCV$ processes and collisional side-feeding combined with
Coster-Kronig processes quickly fill the atomic $L$-shell while
pushing outer levels successively above $V_0$ (cf.\ 
Fig.~\ref{fig:model}). The binding energy of the $2s$ level grows by
several 10~eV with increasing $L$ population. A below-surface
transition involving the ($2s \mapsto 1s$) part should therefore
reflect the range of the $2s$ binding energies convoluted with the VB
electron density. This is in obvious contradiction to the small {\it
  KLV}$_W$ peak width.

Furthermore, at least one additional peak arising from ($2p \mapsto
1s,V_W \uparrow$) transitions should appear because spectral lines on
the high-energy side of the $KLL$ region representing ($2p \mapsto
1s,2p$) transitions give evidence for a partially occupied $2p$ level
below the surface.  In view of the fact that the spectra of metastable
projectiles display a similar shape as those detected for their ground
state counterparts like the Ar$^{8+}$ and Ar$^{9+}$ spectra above the
211-eV peak in Fig.~\ref{fig:argon}, we can additionally conclude that
appreciable inner-shell populations do not occur as the ion descends
towards the surface.  We remark that the prominence of the $KLV_W$
peak in C$^{4+}$ with respect to N$^{5+}$, O$^{6+}$ and Ne$^{8+}$
might be rooted in their low binding energies of the metastable level
possibly favoring high transition rates. Such a tendency has been
found for $LCV$ transitions \cite{Die96} the participating electron
levels of which are separated in a similar way.  At present, a more
precise quantitative treatment for $KLV_W$ and $LMV_W$ processes is
still missing.


With the preceding experimental evidence combined with our
argumentation we supplied convincing proof that the distinct {\it
  KLV}$_W$ peak exclusively occurring in the spectra of metastable
projectiles is generated before bulk penetration.  Provided that a
sufficient statistical quality of the spectra is available, the {\it
  KLV}$_W$ intensity contains valuable information on the interaction
mechanisms and duration of the above-surface phase.  Due to its low
transition rate and the limited interaction time in front of the
surface it sensitively reacts to changes in the perpendicular
projectile velocity.  The discovery of the $KLV_W$ peak therefore sets
a novel benchmark for the scrutiny of previously proposed
above-surface interaction models and opens an exciting perspective for
future research on HCI-surface collisions.  In particular for the very
active field of HCI-insulator scattering, the $KLV_W$ and $LMV_W$ peak
profiles may be exploited to experimentally study resonant charge
transfer, atomic level shifts, projectile trajectories and energy
deposition near the surface in a new manner.


This work was sponsored by the German Bundesministerium f\"ur Bildung,
Wissenschaft, Forschung und Technologie under Contract No. 13N6776/4.
We are also grateful for support from the Ministerium f\"ur
Wissenschaft und Forschung des Landes NRW\@. One of us (JJD) appreciates
fruitful discussions with U.~Thumm and R.~D\'{\i}ez Mui{\~n}o.


\newpage


\begin{references}

\bibitem{Arn97}
A. Arnau, F. Aumayr, P.~M. Echenique, M. Grether, W. Heiland, J. Limburg, R.
  Morgenstern, P. Roncin, S. Schippers, R. Schuch, N. Stolterfoht, P. Varga,
  T.~J.~M. Zouros, and HP. Winter, Surface Science Reports {\bf 27},  113
  (1997).

\bibitem{Mey91}
F.~W. Meyer, S.~H. Overbury, C.~D. Havener, P.~A. {Zeijlmans van Emmichoven},
  and D.~M. Zehner, Phys. Rev. Lett. {\bf 67},  723  (1991).

\bibitem{Gre96}
M. Grether, A. Arnau, R. K\"ohrbr\"uck, A. Spieler, and N. Stolterfoht, Nucl.
  Instrum. Methods Phys. Res. B {\bf 115},  157  (1996).

\bibitem{Tho97}
J. Thomaschewski, J. Bleck-Neuhaus, M. Grether, A. Spieler, D. Niemann, and N.
  Stolterfoht, Nucl. Instrum. Methods Phys Res. B {\bf 125},  163  (1997).

\bibitem{And91A}
H.~J. Andr\"a, A. Simionovici, T. Lamy, A. Brenac, G. Lamboley, S. Andriamonje,
  J.~J. Bonnet, A. Fleury, M. Bonnefoy, M. Chassevent, and A. Pesnelle, Z.
  Phys. D {\bf 21, suppl.},  135  (1991).

\bibitem{Bri96}
J.-P. Briand, B. d'Etat Ban, D. Schneider, M.~A. Briere, V. Decaux, J.~W.
  McDonald, and S. Bardin, Phys. Rev. A {\bf 53},  2194  (1996).

\bibitem{Win96A}
H. Winter, J. Phys: Condens. Matter {\bf 8},  10149  (1996).

\bibitem{Die96}
R. {D\'{\i}ez Mui{\~n}o}, N. Stolterfoht, A. Arnau, A. Salin, and P.~M.
  Echenique, Phys. Rev. Lett. {\bf 76},  4636  (1996).

\bibitem{Lim95a}
J. Limburg, S. Schippers, I. Hughes, R. Hoekstra, R. Morgenstern, S. Hustedt,
  N. Hatke, and W. Heiland, Nucl. Instrum. Methods Phys. Res. B {\bf
  98},  436  (1995).

\bibitem{Koe94}
R. K\"ohrbr\"uck, M. Grether, A. Spieler, N. Stolterfoht, R. Page, A. Saal, and
  J. Bleck-Neuhaus, Phys. Rev. A {\bf 50},  1429  (1994).

\bibitem{Gre95}
M. Grether, A. Spieler, R. K\"ohrbr\"uck, and N. Stolterfoht, Phys. Rev. A {\bf
  52},  426  (1995).

\bibitem{Duc97}  
J. Ducr\'ee, F. Casali, and U. Thumm, Phys. Rev. A {\bf 57}, 338
  (1998).

\bibitem{Sch94}
S. Schippers, J. Limburg, J. Das, R. Hoekstra, and R. Morgenstern, Phys. Rev. A
  {\bf 50},  540  (1994).

\bibitem{Das93}
J. Das and R. Morgenstern, Physical Review A {\bf 47},  R755  (1993).

\bibitem{Duc97A} J. Ducr\'ee, J. Mrogenda, E. Reckels, M. R\"uther, A.
  Heinen, Ch. Vitt, M. Venier, J. Leuker, R.  {D\'{\i}ez Mui\~{n}o},
  and H.~J. Andr\"a, Phys. Rev. A {\bf 57}, 1925 (1998).

\bibitem{Cow81}
R.~D. Cowan, {\em The Theory of Atomic Structure and Spectra \/} (University of
  California Press, Berkeley, 1981).

\bibitem{Bur91}
J. Burgd\"orfer, P. Lerner, and F.~W. Meyer, Phys. Rev. A {\bf 44},  5674
  (1991).

\end{references}

\pagebreak


\begin{table}[htbp]
  \begin{center}
    \leavevmode
    \begin{tabular}{|c|c|c|c|c|} \hline 
      HCI & configuration & $V_W$ [eV] & calc.\ peak energy [eV] & expt.\ peak
      energy [eV] \\ \hline
       C$^{4+}$ &   $1s2s4s^24p^2$ & 4.20 & 294.08 & 294$\pm$1   \\ \hline
       N$^{5+}$ &   $1s2s4s^24p^3$ & 5.25 & 413.26 & 415$\pm$2   \\ \hline
       O$^{6+}$ &   $1s2s5s^25p^4$ & 4.30 & 556.55 & 554$\pm$2   \\ \hline
      Ne$^{8+}$ &   $1s2s5s^25p^6$ & 5.96 & 898.30 & 893$\pm$10  \\ \hline
      Ar$^{8+}$ & $2p^53s6s^26p^6$ & 4.62 & 245.28 & 247$\pm$1   \\ \hline
          \end{tabular}
          \caption{$KLV_W$ and $LMV_W$ peak positions
            as determined by Cowan code simulation and experiment.
            Also listed are the initial configurations entering the
            Cowan code input file and the binding energies of their
            outermost orbital which approximately equal the Al work
            function of $W$=4.25~eV.}
    \label{tab:KLVW}
  \end{center}
\end{table}

\end{document}